\documentclass[preprint]{aastex}

\usepackage{graphicx}

\shorttitle{New nearby M-dwarf companions}
\shortauthors{Kirkpatrick {\it et al.}}

\begin{document}

\title{ Three newly-discovered M-dwarf companions of Solar Neighbourhood stars}

\author{J. Davy Kirkpatrick$^1$}
\affil {Infrared Processing and Analysis Center, 100-22, California Institute
of Technology, Pasadena, CA 91125; e-mail: davy@ipac.caltech.edu}

\author {James Liebert}
\affil {Steward Observatory, University of Arizona, Tucson, AZ 85721; \\
e-mail: liebert@as.arizona.edu}

\author {K. L. Cruz$^2$}
\affil {Dept. of Physics \& Astronomy, University of Pennsylvania, 209 S. 33rd
Street, Philadelphia,  PA 19104-6396; e-mail: kelle@hep.upenn.edu}

\author{J. E. Gizis}
\affil {Infrared Processing and Analysis Center, 100-22, California Institute
of Technology, Pasadena, CA 91125; e-mail: gizis@ipac.caltech.edu}

\author {I. Neill Reid$^2$}
\affil {Dept. of Physics \& Astronomy, University of Pennsylvania, 209 S. 33rd
Street, Philadelphia,  PA 19104-6396 \\
and \\
Space Telescope Science Institute, 3700 San Martin Drive, Baltimore,
MD 21218; \\
e-mail: inr@stsci.edu}

\begin{abstract}
We present low-resolution spectroscopy of newly-discovered 
candidate companions to three stars in the  Solar Neighbourhood.  
All three companions are M dwarfs, with spectral types ranging from M4 to M9.5.
In two cases, G85-55`B' (M6) and G87-9`B' (M4), we have circumstantial
evidence from spectroscopy, photometry and limited
astrometry that the systems are
physical binaries; in the third, G216-7B (M9.5), comparison of POSS II IIIaF
plate material and the 2MASS image indicates common proper motion. 
The primary star in this system, G216-7A (M0), is itself 
an unresolved, nearly equal-mass binary. All three low-mass companions
are highly likely to be stellar in nature, although G216-7B lies
very close to the hydrogen-burning limit.

\end{abstract}


\altaffiltext{1}{Portions of the data presented here were obtained at the W.M.
Keck Observatory, which is operated as a scientific partnership among the
California Institute of Technology, the University of California, and the
National Aeronautics and Space Administration. The Observatory was made
possible by the generous financial support of the W.M. Keck Foundation.
We also make use of data obtained with the Palomar 60-inch telescope, 
owned jointly by the California Institute of Technology and the 
Carnegie Institution of Washington.}

\altaffiltext{2}{Visiting Astronomer at the Infrared Telescope Facility, which is operated 
by the University of Hawaii under contract from the National Aeronautics and Space Administration.}

\section {Introduction}

The constituents of the immediate Solar Neighbourhood provide a snapshot of the
average population of the Galactic Disk. Identifying and characterising those
constituents has been a major focus of stellar research since the early twentieth
century, yet our current catalogues patently remain incomplete, even for
distance limits as small as 10 parsecs. The last few years have seen a renewed emphasis 
on this issue with, in particular, the emergence of the NASA/NSF NStars initiative
(Backman {\sl  et al.}, 2000). As part of that project, the present authors, 
together with other collaborators, are using a variety of techniques to exploit the potential
of the new generation of near-infrared sky surveys, notably 2MASS (Skrutskie {\sl  et al.}, 
1997), in searching for previously-unrecognised nearby stars and brown dwarfs. 
This paper presents some of our first results. 

As with any census, nearby star catalogues are most incomplete for the least
prominent members of the community: some, like brown dwarfs, 
are born to insignificance; others, like white dwarfs, acquire insignificance; still
others, through their proximity to more massive and more luminous companions, 
have insignificance thrust upon them. Here, we describe observations of stars
in the last category: three M dwarfs, each a candidate companion of a Solar
Neighbourhood late-K/early-M dwarf. Section 2 outlines the identification of those
candidates and our spectroscopic observations; section 3 describes the three
systems; and section 4 summarises our conclusions. 

\section {Target identification and observations}

The three M dwarfs discussed in this paper were identified as candidate 
companions of nearby  dwarfs based on photometry from the 2MASS database. 
 All three potential primaries have annual proper motions of $\mu \sim 0.3$ arcsec yr$^{-1}$
and were discovered  in the course of 
the Lowell Observatory proper motion survey (Giclas, Burnham \& Thomas, 1971).
Two of the fainter stars (the candidate
companions to G85-55 and G87-9) were found (by JDK) as part of a larger
survey aimed specifically at searching the environs of known nearby stars
for candidate companions; the third (G216-7B) was uncovered (by KLC) as 
part of an NStars photometric survey for candidate M and L dwarfs within
20 parsecs of the Sun. Table 1 lists  2MASS astrometry and photometry of
each component of these candidate binary systems and Figure 1 plots their
location on the ((J-H), (H-K$_S$) plane. The three brighter stars have
colours consistent with late-type K or early-type M dwarfs, while the
fainter stars lie on the M-dwarf sequence, with G216-7B close to the
M/L boundary.

\subsection {Spectroscopy}

The three candidate companions were observed spectroscopically using the 
Low-Resolution Imaging Spectrograph (LRIS, Oke {\sl  et al.}, 1995) on the
Keck I 10-metre telescope. The instrumental set-up matches 
our previous observations of ultracool dwarfs: we use the 400 line mm$^{-1}$
grating blazed at 8500\AA\ with a 1 arcsecond slit to provide coverage from
6,300 to 10,100\AA\ at a resolution of 9\AA. The spectra were extracted,
wavelength- and flux-calibrated using standard techniques described
by Kirkpatrick {\sl  et al.} (1999), and Figure 2 plots the resultant spectra.

All three stars are clearly M dwarfs. We have determined spectral types using the
TiO5 index defined by Reid {\sl  et al.} (1995), 
\begin{displaymath}
{\rm TiO5} \ = \ {{\langle F(7126-7135) \rangle} \over {\langle F(7042-7046) \rangle}}
\end{displaymath}
measuring the overall depth of the 7050\AA\ TiO bandhead. As shown in their Figure 2,
this index has a bimodal distribution with spectral type, decreasing 
in a near-linear manner from TiO5$\sim0.9$ at K7 to a minimum value of $\sim0.15$ at M6.5,
but increasing at later types as VO absorption erodes the peak of the bandhead.
Considering the three M dwarfs observed here,  G87-9`B' has TiO5=0.39, corresponding to 
spectral type M4. G85-55`B' is of later spectral type, with the TiO5 ratio of 0.23 consistent 
with spectral type M6 or M7.5; based on the appearance of the VO bands at 7000
and 7400\AA\, the earlier type is more appropriate. G216-7B is the latest of the
three dwarfs, with strong VO and TiO5=0.51, indicating a spectral type of M9.5. 
The two later-type dwarfs exhibit H$\alpha$ in emission, with equivalent widths of 9\AA\
in G85-55`B' and 0.7\AA\ in G216-7. Visual inspection of the spectra (by JDK) results in 
nearly identical spectral types (JDK estimates M6.5 for G85-55`B').
We have also checked for the 6708\AA\ absorption line
due to neutral lithium; none of the three dwarfs show detectable absorption, with an
upper limit on the equivalent width of 0.3\AA\ for G87-9`B' and $<3 \AA\ $ for the 
two cooler dwarfs. 

\subsection {Near-infrared astrometry}

As discussed further below, the 2MASS scans provide the earliest-epoch imaging of
G85-55'B' and G87-9'B'. We therefore supplemented these data with K-band images of
both systems, obtained on January 19, 2001 using NSFCAM on the NASA Infrared Telescope
Facility. Conditions were photometric, with 0.7 arcsecond seeing. We used the
0.148 arcsec pix$^{-1}$ platescale, taking images in a five-point dither pattern,
with each image the sum of $25 \times 0.08$ second exposures. This configuration
provided unsaturated images of both the primary star and the potential companion,
allowing accurate measurement of their relative positions.

\section {The systems}

The individual stars in each system are discussed in more detail in this section. Table 2
collects published photometric and astrometric observations of the potential
primaries. Table 3 gives the derived luminosities, masses and level of 
chromospheric or coronal activity.

\subsection {G85-55}

The brighter star in this system is also known as HD 248184, BD +19 872, LTT 11613 and
GJ 3338. As Table 2 shows, this is the bluest of the three potential primaries. 
Although the SIMBAD database lists a spectral type of K5, there is no recent
published spectroscopy and the optical/near-infrared colours suggest that the
spectral type is closer to K4. Heintz (1994) has determined an absolute trigonometric
parallax of $42\pm6$ milliarcseconds ({\it mas}). As Figure 3 shows, this 
places G85-55`A' slightly below the (M$_V$, (V-K$_S$) main sequence, although
Sandage \& Kowal's (1986) UBV measurements indicate near-solar metal abundance. 
Figure 4 shows similar agreement in the (M$_J$, (J-K$_S$) plane. 

G85-55`B' lies only 9.4 arcseconds from G85-55`A', and is not  detectable on
either POSS I or POSS II.  The 2MASS observations, made on 3 November 1997, therefore
provide the first epoch astrometry of this system. Our IRTF observations provide
a baseline on 3.2 years, allowing a first test for common proper motion
\footnote{ With a projected separation of $\sim225$AU, the binary system would
have a period exceeding 2000 years, and orbital motion has a negligible effect
on the relative proper motions. The same circumstances hold for the other two
binaries considered in this paper.} . The
relative offset between the two stars (`A' to `B') in the 2MASS observations
is $\Delta \alpha = 2.8\pm0.44$ arcseconds, $\Delta \delta = 9.0\pm0.20$ arcseconds; the proper
motion of the primary is 0.31 arcsec yr$^{-1}$, $\theta = 131^o$, corresponding
to relative motion of +0.265 arcsec in Right Ascension and -1.23 arcsecond in Declination
if `B' is stationary with respect to `A'; our measured offsets from the IRTF data
are  $\Delta \alpha = 3.1\pm0.25$ arcseconds, $\Delta \delta = 8.8\pm0.25$ arcseconds.
With such a short time baseline, these data do not provide conclusive evidence, but
the observations are at least consistent with common proper motion between the 
two stars.

We can check whether the spectroscopic parallax of the companion is
consistent with  the near-infrared photometric properties.
If we assume a parallax of $42\pm6$ {\it mas}, the 
2MASS photometry implies M$_K(B) = 9.7\pm0.3$. In comparison, the nearby M5.5 dwarfs, Gl 65A
and GJ 1116A have M$_K$=8.78 and 8.83; 
the M6 dwarfs Gl 406 (Wolf 359) and Gl 65B (UV Ceti) have
M$_K=9.18$ and M$_K=9.17$; and the M6.5 dwarfs LHS 3003 and GJ 1111 have M$_K$=9.96 and
9.46; thus, an association with G85-55`A' is broadly consistent with the observed spectral type.
Placing G85-55'B' at 42 parsecs also gives good agreement with the main sequence in 
the (M$_J$, (J-K$_S$)) plane, as illustrated in Figure 4.

While statistical arguments are always vulnerable when considering one object, we can
also consider the probability of random association. Given the (M$_K$, spectral type)
results listed above, we 
assume that our spectral type estimate, M6$\pm0.5$, corresponds to M$_K=9.2\pm0.5$. In that case,
 we derive a spectroscopic distance of 
$r_{spec} = 39.8^{+10.3}_{-8.2}$ parsecs and place the star within a spherical
 shell of volume $\sim4 \times 10^5$ pc$^3$.
There are 13 dwarfs with spectral types between M5.5 and M6.5 in the northern 8-parsec
sample (Reid \& Hawley, 2000: RH2000), corresponding to a space density of $\rho_{M6} =0.008$ stars pc$^{-3}$.
Our sampling shell therefore contributes a surface density of 0.078 M6 dwarfs sq.deg.$^{-1}$
with the appropriate apparent magnitude. There is
a corresponding probability of $1.9 \times 10^{-6}$ of finding an M6 dwarf within
10 arcseconds of a randomly-chosen position on the sky. While not conclusive, these result
are strongly suggestive of a physical association between the K5
dwarf, G85-55`A', and the M6 dwarf, G85-55`B'. If so, the physical separation is $\sim225$
astronomical units. 

Considering G85-55`B' itself, the spectral type of M6 implies an effective temperature
of $2800\pm150$K (Leggett {\sl et al.}, 1996). Figure 5 matches those limits against
theoretical tracks from Burrows {\sl  et al.} (1993, 1997) and Baraffe {\sl  et al.} (1998).
Briefly comparing those model predictions, the former tracks (Arizona models) predict
higher temperatures (at a given age) than the latter (Lyon) tracks for masses above 
$\sim0.08 M_\odot$, but lower temperatures for masses below $\sim0.07 M_\odot$. The 
hydrogen-burning limit lies at a slightly lower mass in the Lyon models, with a 0.075$M_\odot$
Lyon dwarf falling above the limit, while a 0.075M$_\odot$ Arizona dwarf is a transition
object. Considering G85-55'B', 
our spectroscopy failed to detect Li I 6708\AA, but  sets an upper limit
of only 3\AA\ equivalent width, while the expected equivalent width for no depletion at these
temperatures is 1 to 2 \AA. However, our temperature estimates allow us to set upper limits 
on the mass of G85-55`B': if we adopt the Arizona models, we infer
$M \le 0.1 M_\odot$; in the case of the Lyon models, $M \le 0.11 M_\odot$. 

We can refine the mass estimate by considering the level of chromospheric
activity, measured by the flux ratio, $F_\alpha \over F_{bol}$; youthful dwarfs
can be expected to have above average activity, and Figure 5 shows that we require
$\tau < 0.25$ Gyrs for $M < 0.075 M_\odot$.  We can derive the H$\alpha$ line flux directly
from our spectrum as F$_\alpha = 3.5 \times 10^{-15} \ {erg \ cm^{-2} \ s^{-1}}$.
Previous experience has shown that the J-band bolometric correction provides a
robust method of estimating luminosities in M and L dwarfs (Reid {\sl et al.}, 2001). 
Leggett {\sl et al.} (1996) derive BC$_J$=1.99 mag. for the M5.5/M6 binary Gl 65AB and
BC$_J$=2.06 mag. for the M6.5 dwarf GJ 1111. Given these results, we adopt
BC$_J = 2.0$ mag. for G85-55`B' and calculate m$_{bol} = 14.5$. Taking $M_{Bol}(\odot) = 4.75$,
we find a ratio of ${F_\alpha \over F_{bol}} \ = \ 8.6 \times 10^{-5}$ (Table 3). 
This is $\sim10\%$ below the
average activity level of M6 dwarfs in both the Hyades cluster (Reid \& Mahoney, 2000) and
the general field (Gizis {\sl et al.}, 2000). This suggests that G85-55`B' has an age
$\tau > 0.6$ Gyrs and a mass exceeding 0.08 $M_\odot$.

\subsection {G87-9}

The supposed primary of this system is also known as G103-65, LP 254-40, GJ 3411
and HIP 32723. 
Reid {\sl et al.} (1995) derive a spectral type of K5 based on a TiO5 index of 0.94. This
is consistent with the optical/near-infrared colours listed in Table 2. The Hipparcos
parallax indicates a distance of $27.6\pm1.3$ parsecs and M$_V=8.4\pm0.1$.
As Figures 3 and 4 show, this places the star close to the main sequence. 

G87-9`B' lies only 6.5 arcseconds from G87-9`A' so, as with G85-55, 
the 2MASS observation, made on 23 November 1998, provides first epoch relative astrometry. 
We measure $\Delta \alpha = +2.5\pm0.3$ arcseconds (`A' to `B') and $\Delta \delta = -6\pm0.3$
arcseconds.  Our IRTF observations provide a baseline of only 2.145 years, during which time
the primary ($\mu = 0.32$ arcsec. yr$^{-1}$, $\theta = 153^o$) is
predicted to move 0.31 arcseconds East and 0.6 arcseconds South; we measure offsets
of $\Delta \alpha = +3.1\pm0.25$ arcseconds and $\Delta \delta = -5.8\pm0.25$ arcseconds.
As with G85-55, the recent measurements are consistent with common proper motion, but
do not set strong constraints on the relative motion.

If we assume that the two stars are associated, then the
absolute magnitude is M$_K$(B)=8.0. The (M$_J$, (J-K$_S$) colour-magnitude
diagram offers no discriminatory power for mid-type M dwarfs, but we can compare
the absolute magnitude for consistency against the measured spectral type, M4$\pm0.5$.
There are fifteen M4 dwarfs in the northern 8-parsec sample (RH2000),
eleven of which have K-band photometry; the average absolute magnitude is
$\langle M_K \rangle = 7.50$, with an rms dispersion of $\sigma_K = \pm 0.46$ 
magnitudes. In comparison, we derive $\langle M_K \rangle = 7.80$,  $\sigma_K = \pm 0.69$,
from K-band data for nine of the 13 M4.5 dwarfs listed by RH2000, and
 $\langle M_K \rangle = 7.27$,  $\sigma_K = \pm 0.61$ from observations of nine of the 
15 M3.5 dwarfs. While the constraints are relatively weak, the inferred absolute 
magnitude of G87-9`B' for $r = 27.6$ parsecs is consistent with the observed spectral type.

As with G85-55, we can calculate the probability of a chance association. Given
the ($\langle M_K \rangle$, spectral type) values listed above, we take our spectral type
estimate as corresponding to limits of $6.8 < M_K < 8.1$. These place a
K$_S$=9.8 magnitude M4 dwarf at a distance between 39.8 and 21.9 parsecs, within a
spherical shell of volume $2.2 \times 10^5$ parsecs. There are 45 dwarfs in the RH2000
8-parsec sample with spectral types between M3.5 and M4.5, giving a space density 
$\rho_{M4} = 0.028 \ {\rm stars \ pc^{-3}}$. This implies an expected surface density of 
0.149 stars sq.deg.$^{-1}$, and a probability of $\sim 1.8 \times 10^{-6}$ of
finding an M4 dwarf at the appropriate apparent magnitude within 7 arcseconds of a
randomly chosen point. 

Again, the available constraints favour the identification of G87-9`B' as a
binary companion of G87-9`A' at a physical separation of 200 AU. Given the
spectral type and the absence of both lithium absorption and H$\alpha$ emission, 
G87-9`B' is likely to be a disk dwarf with $\tau > 1$ Gyr. and a mass of $\sim0.18 M_\odot$.

\subsection {G216-7}

This is the best studied of the three systems and the only confirmed binary.
The primary is also known as G189-30, BD+38 4818, LTT 16634, HIP 111685 and GJ 4287.
Reid {\sl et al.} (1995) measure a spectral type of K7 based on a TiO5 index of 0.81. 
Our LRIS spectrum of the secondary provides an opportunity for an independent
measurement through the presence of scattered light from G216-7A; we
measure TiO5=0.75, corresponding to a spectral type of M0. Weis (1987) has
obtained VR$_K$I$_K$ photometry, and those colours are more consistent with
the later spectral type. 

G216-7A was observed by Hipparcos (ESA, 1997) and lies at a distance 
of $18.89^{+0.72}_{-0.65}$ parsecs, within the 25 parsec limit of the NStars project. 
At that distance modulus, the star lies significantly above the main sequence in
the (M$_V$, (V-K$_S$) plane (Figure 3), and, even allowing for the parallax uncertainties, 
is brighter than the earlier-type G87-9`A' in both M$_V$ and M$_J$ (Figure 4). In fact,
Hipparcos has resolved the star as a binary with separation $0.144\pm0.019$ arcseconds
(PA = 355$^o$, epoch 1991.25), with a magnitude difference $\Delta H_P = 0.44\pm0.96$
magnitude.  If we assume a V-band 
flux excess of 0.5 magnitudes above the single-star main-sequence, 
then the two components have absolute magnitudes
of M$_V{\rm (Aa)} \sim 8.5$ and  M$_V{\rm (Ab)} \sim 9.1$. 

Given the relatively small magnitude difference, G216-7A might be resolved as
a double-lined spectroscopic binary, while a close binary system might be
expected to have unusually high chromospheric and coronal activity.
Gizis, Reid \& Hawley (in prep.) have a 
single high-resolution echelle spectrum, taken using McCarthy's spectrograph on the
Palomar 60-inch telescope, but there is no evidence for line doubling.  
Those data do permit measurement of the  radial velocity, 
V$_r = -53.1\pm1.5 {\rm \ km \ s^{-1}}$.
Combined with the Hipparcos astrometry, this gives heliocentric space motions (U, V, W)
of (20.8, -56.6, -11.1), where U is positive towards the Galactic Centre. These
motions are consistent with membership of the old disk population.

The echelle spectrum also provides a measure of chromospheric activity: 
there is no detectable Balmer
emission, with H$\alpha$ absorption with equivalent width 0.69\AA\
(comparable to Hyades M0 dwarfs); emission is present
at Ca II H and K, with equivalent widths of 3.71 and 2.45 \AA\ respectively, but 
activity at this level is not unusual for late-K/early-M dwarfs. Huensch {\sl et al.} (1999)
derive an X-ray luminosity of $7.7 \times 10^{27} {\rm \ erg \ s^{-1}}$ from
ROSAT observations. Combined with our estimate of M$_{bol}$ (Table 3), 
we derive $\log{L_X \over L_{bol}} = -4.88$, a low level of coronal activity
for early-type M dwarfs. 

 Turning to the companion, G216-7B lies 33.6 arcseconds from the primary
and, at that distance, is clearly visible on the POSS II IIIaF survey plate; indeed,
the companion may be barely visible on the POSS I E plate (Figure 6). In any case,
the time baseline between the POSS II F plate (3 September 1989) and the 2MASS
observation (10 October 1998) is sufficient to show that the position angle and
separation between G216-7A and G216-7B remains unchanged, despite motion of almost 3
arcseconds, confirming G 216-7B as a physical companion at a separation of 635 AU. 

While we have classed G216-7B as spectral class M9.5, it clearly lies very close to
the L dwarf r\'egime, as Figures 1 and 4 illustrate.
With M$_J$=11.97 and M$_K$=10.77, it is over half a magnitude
fainter than the archetypical M9.5 dwarf, BRI0021, and within 0.1 magnitudes of
the absolute magnitude of the L0.5, 2M0746+20 (making due allowance for the fact that
2M0746 is an equal-mass binary). Like most ultracool dwarfs, G216-7B exhibits a low
level of chromospheric activity. The weak H$\alpha$ emission, equivalent width 0.7\AA, corresponds
to a line flux of F$_\alpha = 4.2 \times 10^{-17} \ {\rm erg \ s^{-1}}$. Assuming BC$_J \sim 2$
magnitudes, we derive m$_{bol} = 15.35$, giving ${F_\alpha \over F_{bol}} = 2.5 \times 10^{-6}$.

Based on the spectral type, we estimate the effective temperature as $2100\pm150$K.
Those limits are superimposed on the theoretical tracks plotted in Figure 5. As with
G85-55`B', our non-detection of lithium sets only weak constraints on the mass, since the
expected equivalent width is only 0.5 to 1\AA, as in the M9.5 brown
dwarf, LP 944-20 (Tinney, 1998). However, the inactivity
of the M0 companion suggests an age at least comparable to that of the Hyades, and
probably exceeding 1 Gyr. Under those circumstances, G216-7B is likely to have a mass in
the range 0.065 to 0.08M$_\odot$, regardless of whether one uses the Arizona or
Lyon models as the reference. If $\tau > 1.25$ Gyrs, G216-7B is a transition object or a
star, rather than a brown dwarf.

\section {Summary and conclusions}

We have presented spectroscopic observations of three candidate low-mass
companions to known nearby stars. Each was identified from the 2MASS catalogue
based on their near-infrared colours. All three prove to be M dwarfs with
spectral types M4, M6 and M9.5. None has detectable lithium absorption and,
based on their characteristics (luminosity, colours, activity) and the 
characteristics of the potential primaries, all three are almost certainly stars
rather than brown dwarfs. However, the M9.5 dwarf, G216-7B, lies close to
the hydrogen-burning limit. 

Based on our spectroscopy, together with literature astrometric and photometric
data, two of the three systems, G85-55`A'/`B' and G87-9`A'/`B', can only
be classed as probably binaries; the third, G216-7A/B, is confirmed as 
a common proper motion system based on comparison of POSS II plate material
and 2MASS images spanning a baseline of 9.1 years. With annual proper motions of
$\sim0.3$ arcsec yr$^{-1}$, similar confirming observations of the probable systems
will be possible in the near future.

The available data indicate distances of $24\pm3.5$ parsecs for G85-55, $27.6\pm 1.4$ parsecs for
G87-9 and $18.9\pm0.7$ parsecs for G216-7, all based on trigonometric
parallax measuremements. In each case, the distance
is derived from observations of the (hypothetical) primary. We note that, while all
three are included in the latest version of the Nearby Star Catalogue (as indicated by
the GJ designation), only G216-7 is unequivocally within 25 parsecs of the Sun
and therefore eligible for inclusion in the NStars Database. 

These three candidate binaries were uncovered in the first stages of our
search through the 2MASS database for unrecognised members of the Solar Neighbourhood.
At present, the parent samples are not defined with sufficient precision,
nor do we have sufficient examples in hand,  to permit a reliable estimate of
how many similar systems remain to discovered. 
Nonetheless, it is likely that further examples will be discovered, particularly by
the targetted wide-companion search (the van Biesbroeck project).

Finally, if confirmed, all three of these systems have component separations
exceeding 150 AU. Elsewhere, we have discussed a possible
correlation between the maximum separation of binary components and the 
total system mass (Reid {\sl et al.}, 2001).
The systems discussed here include a K4+M6 (0.7+0.1 $M_\odot$)
binary at $\Delta = 225$AU, a K5+M4 (0.65+0.18 $M_\odot$) binary at $\Delta \sim 200$AU
and an (M0+M0)+M9.5 (0.6+0.6+0.08) triple at $\Delta \sim 635$ AU. All
three fall within the hypothetical ($log(\Delta), M_{tot}$) limits outlined in
Figure 11 of Reid {\sl et al.} (2001).

\acknowledgements 
This NStars research was supported by a grant awarded as part of the
NASA Space Interferometry Mission Science Program, administered by the 
Jet Propulsion Laboratory, Pasadena. We would like to thank Hartmut Jahreiss for
useful comments, particularly pointing out the separate listings of G87-9
on SIMBAD.  INR and JL also acknowledge partial support through a NASA/JPL grant to
2MASS Core Science. JDK acknowledges the support of the Jet Propulsion Laboratory,
California Institute of Technology, which is operated under contract with the
National Aeronautics and Space Administration. This publication makes use of
data from the 2-Micron All-Sky Survey, a joint project of the University of
Massachusetts and the Infrared Processing and Analysis Center, funded by the 
National Aeronautics and Space Administration and the National Science Foundation. 
Our analysis also relies partly on
photographic plates obtained at the Palomar
Observatory 48-inch Oschin Telescope for the Second Palomar
Observatory Sky Survey which was funded by the Eastman Kodak
Company, the National Geographic Society, the Samuel Oschin
Foundation, the Alfred Sloan Foundation, the National Science
Foundation grants AST84-08225, AST87-19465, AST90-23115 and
AST93-18984,  and the National Aeronautics and Space Administration 
grants NGL 05002140 and NAGW 1710.
This research also made use of the SIMBAD database, operated at CDS, Strasbourg, France.

\newpage

\figcaption { The ((J-H), (H-K$_S$)) diagram: crosses are data for nearby stars, spectral
types A to M; L dwarfs are plotted as open squares; the T dwarfs Gl 229B and Gl 570D are
plotted as five-point stars. The extremely red M dwarf is 2M0149+20 (M9.5 pec.).
The six points with errorbars plot the 2MASS colours for  the six stars discussed in
this paper: open circles for G85-55`A'/`B'; open triangles for G87-9`A'/`B'; and
solid points for G216-7A/B. }

\figcaption{Keck LRIS spectra of the three M-dwarf companions discussed in the text. The
spectra have been sky-subtracted, but not corrected for telluric absorption.}

\figcaption { The (M$_V$, (V-K$_S$)) colour-magnitude diagram: the three primary stars
are plotted, using the same symbols 
as in Figure 1. Note that G 216-7A, the reddest object, lies significantly above the main
sequence at its colour ((V-K$_S$)=3.55).}

\figcaption{The location of the six stars discussed in this paper in
the (M$_J$, (J-K$_S$)) colour-magnitude diagram; the symbols have the same meaning
as in Figure 1.}

\figcaption {A comparison between observed parameters for the two cool M dwarfs, G85-55`B' and G216-7B,
and theoretical models. The upper panel plots (T$_{eff}$, $\tau$) evolutionary tracks calculated
by Burrows {\sl et al.} (1994, 1997) for a range of masses (0.02, 0.03, 0.04, 0.05, 0.06, 0.07, 0.075,
0.08, 0.09, 0.10 and 0.11 M$_\odot$), with the heavy line indicating the approximate location of
lithium depletion to 1\% primordial abundance. The horizontally hatched regions mark the 
observational locations of G85-55`B' ($2800\pm150$K) and G216-7B ($2100\pm150$K); both have
depleted lithium. The lower panel plovides the same comparison against the Baraffe {\sl et al.} (1998)
models (masses of 0.02, 0.025, 0.04, 0.05, 0.06, 0.07, 0.072, 0.075, 0.08, 0.09, 0.10 and 0.11 M$_\odot$.}

\figcaption{ The G216-7 system: the upper left and upper right figures plot $5 \times 5$
arcminute regions from the  POSS I E  and POSS II F plates; the lower images plot the 2MASS 
J (left) and K$_S$ data on the same scale. North is up and east to the left.
The arrow indicates the location of G216-7B on each image.}

\newpage
\null

\newpage

\begin{table}
\begin{center}
{\bf Table 1} \\
{ 2MASS astrometry and  photometry}
\begin{tabular}{llrrrr}
\tableline\tableline
Name & 2MASSI & J & H & K$_S$ & (J-K$_S$) \\
\tableline
G85-55`A' & J0514169+195258 & $7.40\pm0.01$ & $6.88\pm0.03$ & $6.74\pm0.01$ & $0.66\pm0.01$ \\
G85-55`B' & J0514171+195307 & $12.53\pm0.07$ & $11.90\pm0.05$ & $11.58\pm0.03$ & $0.95\pm0.08$ \\
G87-9`A' & J0649358+350826 & $7.68\pm0.01$ & $7.09\pm0.02$ & $6.90\pm0.01$ & $0.78\pm0.02$ \\
G87-9`B' & J0649360+350820 & $10.61\pm0.03$ & $10.10\pm0.08$ & $9.76\pm0.03$ & $0.85\pm0.04$ \\
G216-7A & J2237298+392251 & $6.63\pm0.01$ & $6.04\pm0.03$ & $5.86\pm0.02$ & $0.77\pm0.02$ \\
G216-7B & J2237325+392239 & $13.35\pm0.03$ & $12.68\pm0.03$ & $12.15\pm0.03$ & $1.20\pm0.04$ \\

\tableline\tableline
\end{tabular}
\end{center}
\end{table}

\clearpage

\begin{table}
\begin{center}
{\bf Table 2} \\
{Published photometry and astrometry}
\begin{tabular}{llrrrrrrcccl}
\tableline\tableline
Name & V & (U-B) & (B-V) & (V-R)$_C$ & (V-I)$_C$ & (V-K$_S$) &$\mu$ & $\theta$ &$\pi$ & sources \\
      &  &       &       &           &           & &`` yr$^{-1}$ & $^o$ & {\it mas} \\
\tableline
G85-55`A' & 9.47 & 1.11 & 1.17 &  & & 2.73 &0.31 & 131 & $42\pm6$ & 1, 3, 4 \\
G87-9`A' & 10.18 & 1.27 & 1.33 &0.79  &1.54  & 3.28&  0.348 & 132 & $36.16\pm1.82$ & 1, 2, 5 \\
G216-7A & 9.41 &  &  &  0.83 & 1.68 &3.55& 0.341 & 176.4 & $52.94\pm1.94$ & 2, 5 \\
\tableline\tableline
\end{tabular}
\end{center}
Reference sources: 1. Sandage \& Kowal, 1986: UBV \\ 
2. Weis, 1987: VRI (transformed to Kron-Cousins following Bessell \& Weis (1987) \\
3. Giclas, Burnham \& Thomas, 1965: $\mu$, $\theta$ \\
4. Heintz, 1994: $\pi$ \\
5. ESA, 1997: $\mu$, $\theta$, $\pi$ 
\end{table}

\clearpage

\begin{table}
\begin{center}
{\bf Table 3} \\
{Luminosities, Masses and Activity}
\begin{tabular}{lrrrcccccc}
\tableline\tableline
Name & M$_V$ & M$_J$ & M$_{bol}$ & Src. & Mass & Src. & log($F_\alpha \over F_{bol}$) &
log($L_X \over L_{bol}$) \\
\tableline
G85-55`A' & $7.58\pm0.25$ & 5.5 & 7.1 & 1 & 0.70 & 1 & \\
G85-55`B' &     & 10.65 & 12.6 & 2 & 0.07 to 0.10 & 2 & -4.07 & \\ 
G87-9`A' & $8.41\pm0.11$ & 5.91 & 7.6 & 1 & 0.65 & 1 & \\   
G87-9`B' &      & 8.84 & 10.6 & 2 & 0.18 & 3 & \\
G216-7A & $8.03\pm0.07$ & 5.25 & 6.8 & 3 & 0.6+0.6 & 1 & & -4.88 \\
G216-7B &    & 11.97 & 14.0 & 2 & 0.06 to 0.08 & 2 & -5.60 \\
\tableline\tableline
\end{tabular}
\end{center}
Reference sources: \\
Bolometric magnitudes: 1. Flower (1996), ((B-V), BC$_V$) \\
2. BC$_J$ data from Leggett {\sl et al.} (1996) and Reid {\sl et al.} (2001) \\
3. (BC$_I$, (V-I)) relation from Reid \& Mahoney (2000) \\
Masses: 1. Henry \& McCarthy (1993), (M$_V$, log(mass)) relation \\
2. Figure 5, this paper \\
3. Henry \& McCarthy (1993), (M$_K$, log(mass) relation
\end{table}

\setcounter{figure} {0}

\begin{figure}
\plotone{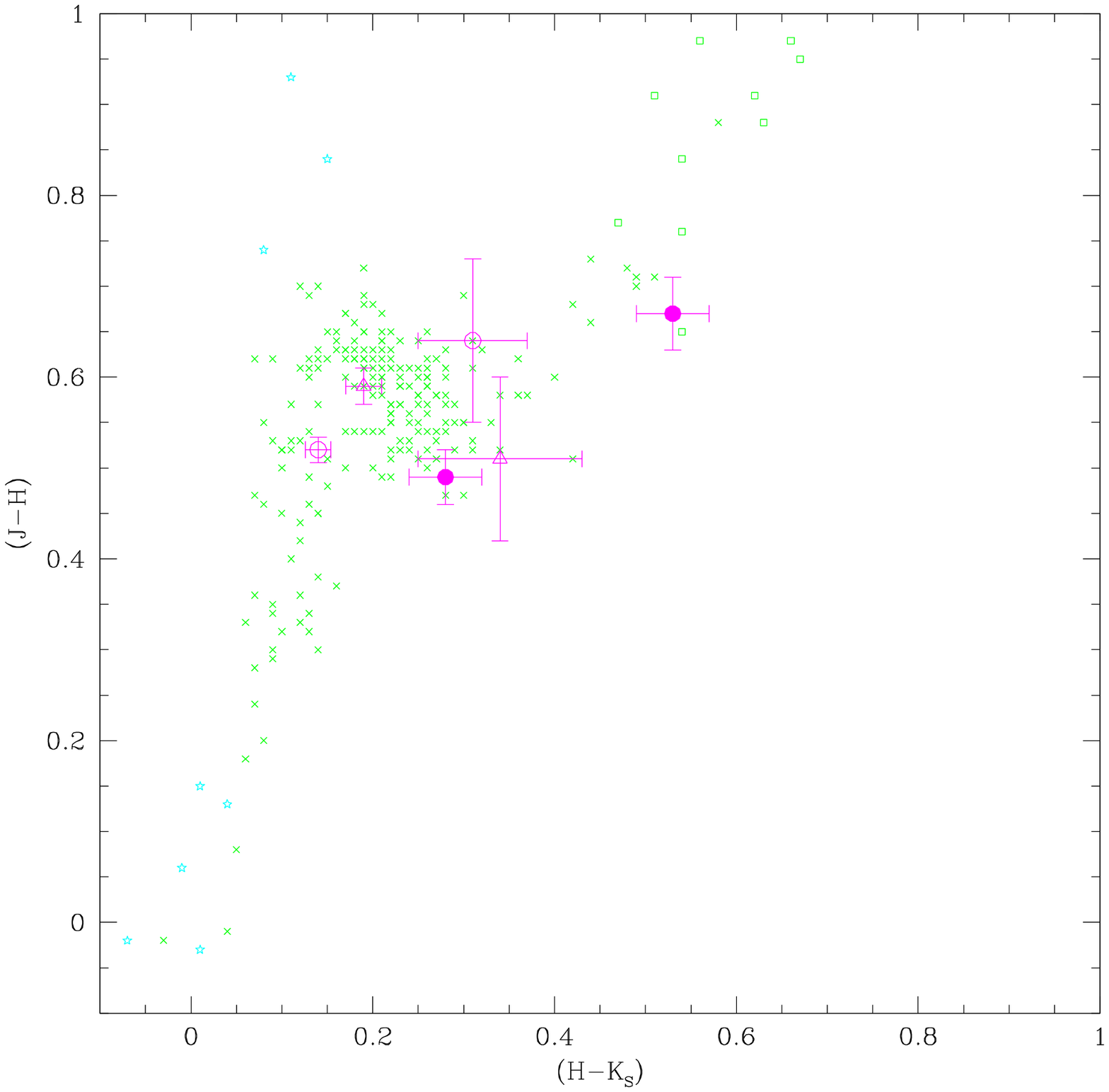}
\caption{}
\end{figure}

\begin{figure}
\plotone{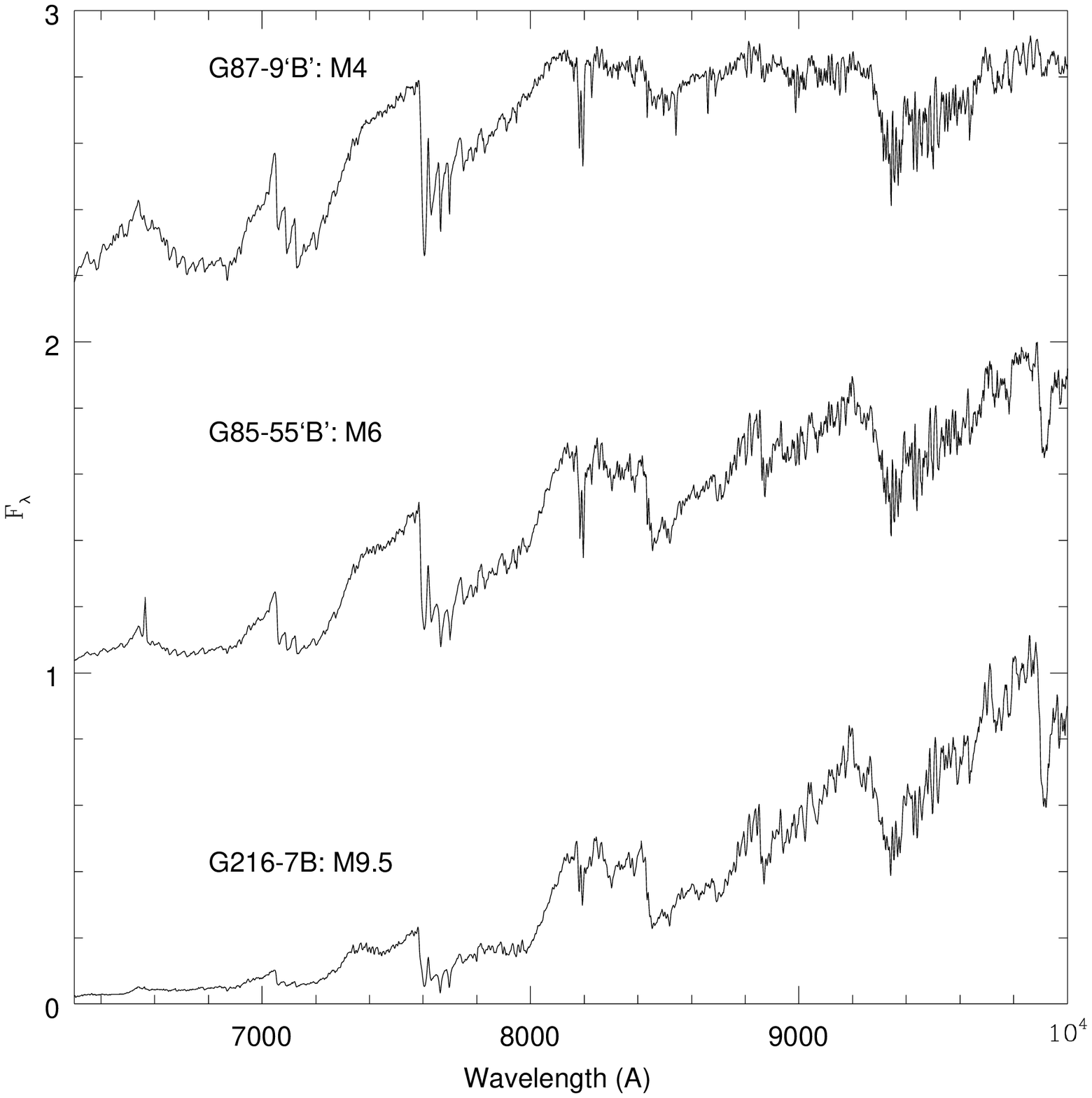}
\caption{}
\end{figure}

\begin{figure}
\plotone{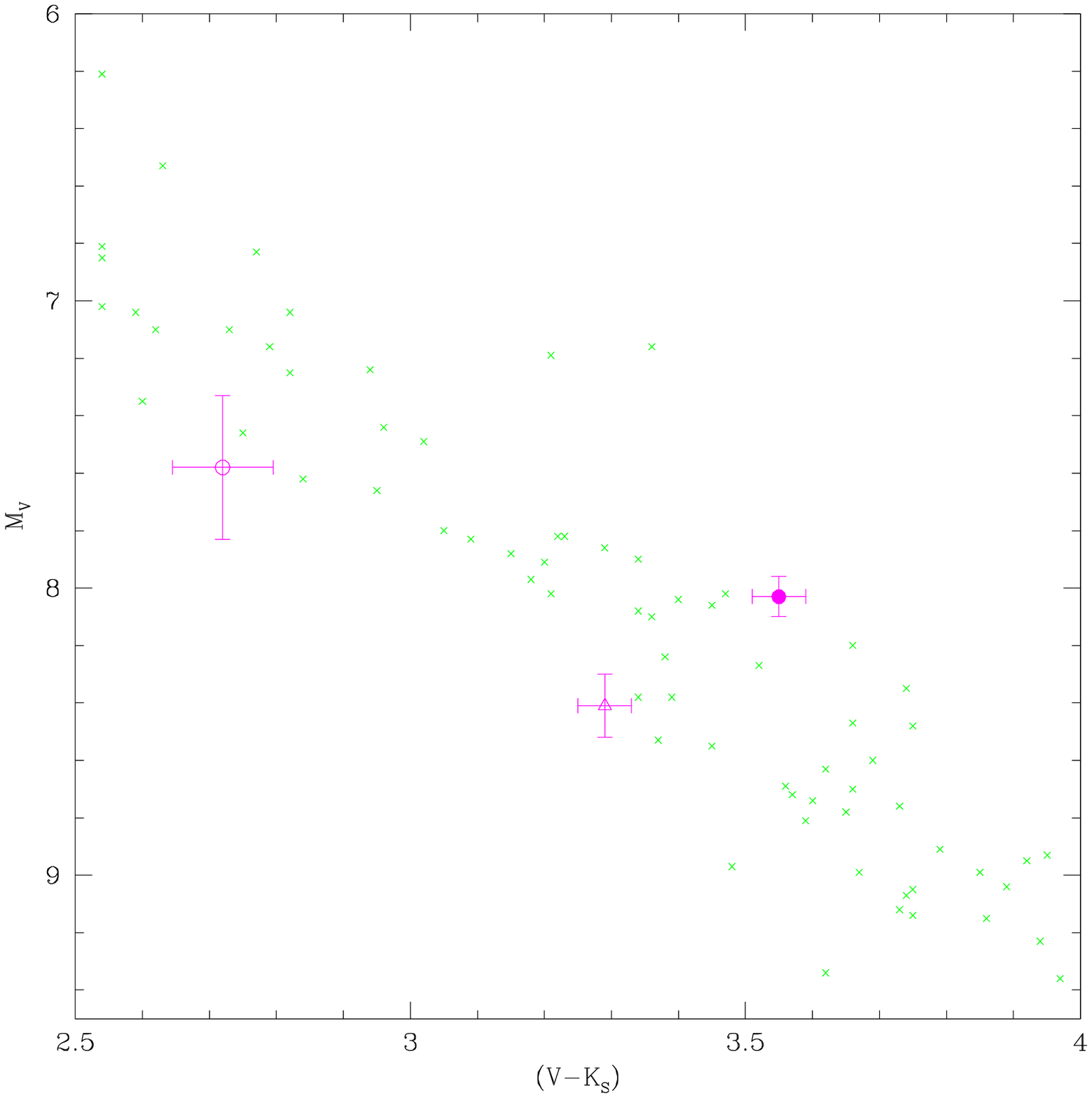}
\caption{}
\end{figure}

\begin{figure}
\plotone{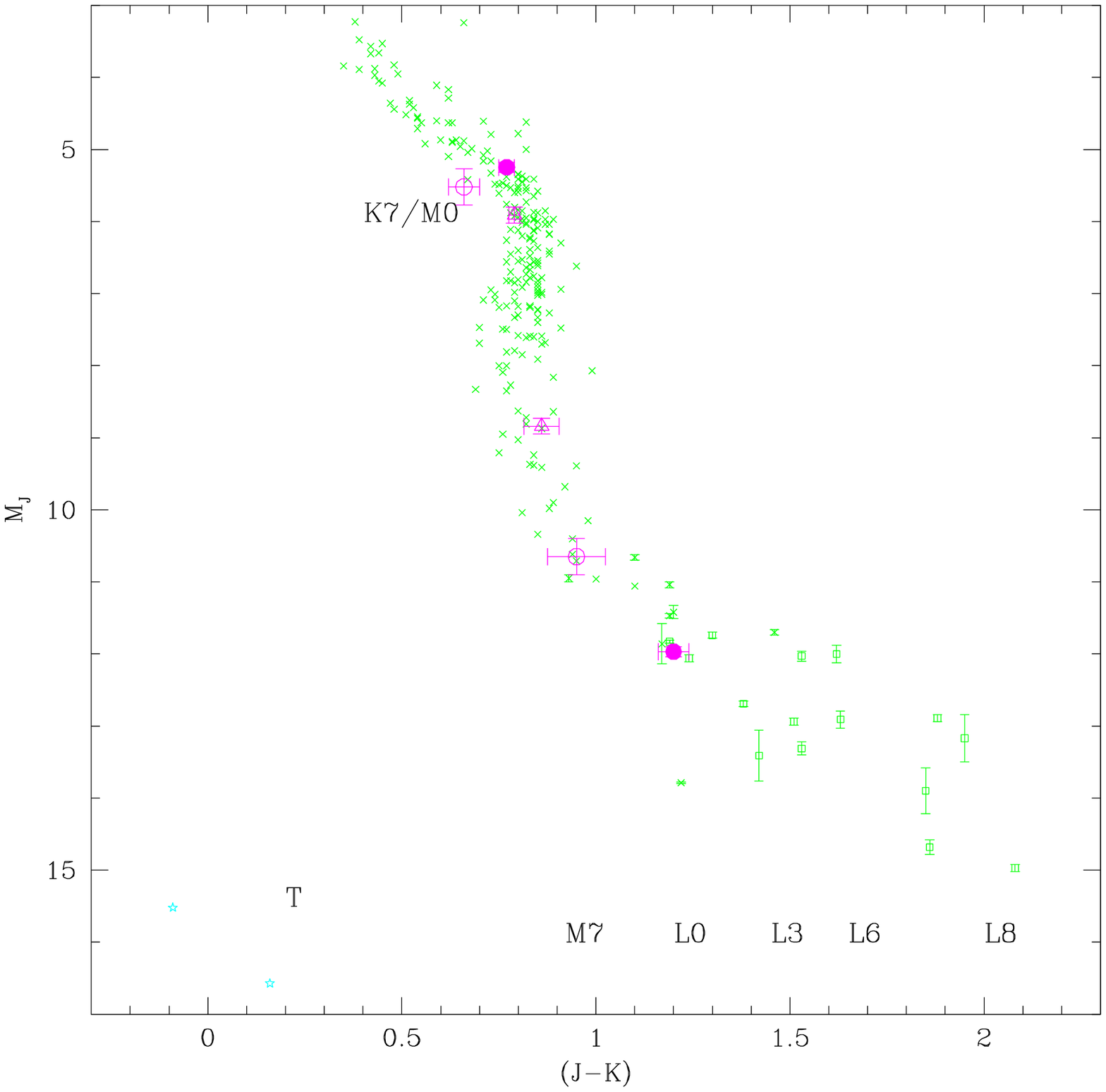}
\caption{}
\end{figure}

\begin{figure}
\plotone{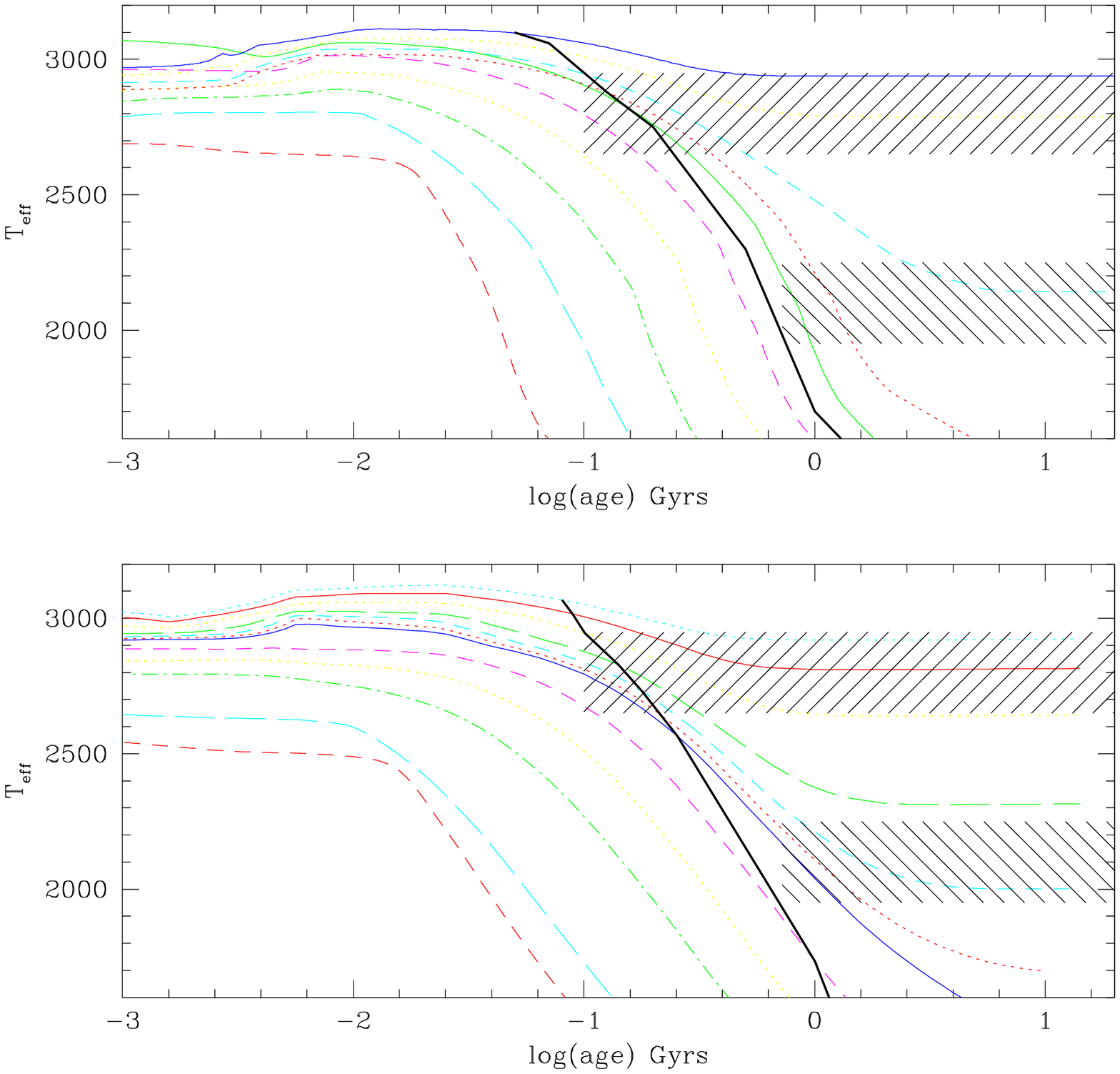}
\caption{}
\end{figure}

\begin{figure}
\plotone{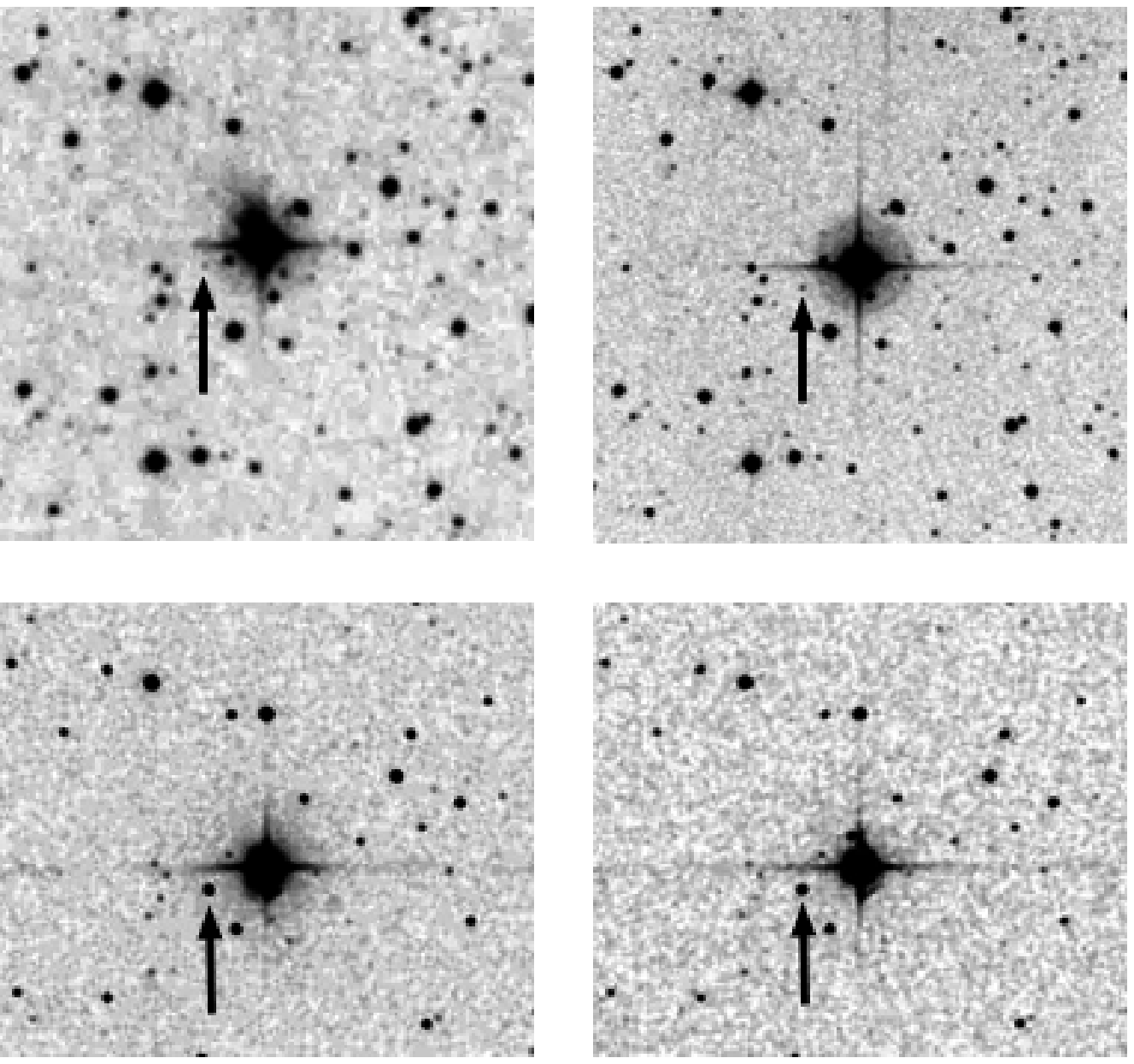}
\caption{}
\end{figure}

\end{document}